\documentclass[12pt,preprint]{aastex}

\def\gtorder{\mathrel{\raise.3ex\hbox{$>$}\mkern-14mu
             \lower0.6ex\hbox{$\sim$}}}
\def\ltsima{$\; \buildrel < \over \sim \;$}
\def\simlt{\lower.5ex\hbox{\ltsima}}
\def\gtsima{$\; \buildrel > \over \sim \;$}
\def\simgt{\lower.5ex\hbox{\gtsima}}

\begin{document}

\title{Science with a wide-field UV transient explorer}

\author{I. Sagiv}
\email{ilan.sagiv@weizmann.ac.il}

\author{A. Gal-Yam\altaffilmark{1}}
\author{E. O. Ofek}
\author{E. Waxman}
\author{O. Aharonson}

\affil{Department of Particle Physics and Astrophysics, Faculty of Physics, The Weizmann
Institute of Science, Rehovot 76100, Israel}

\author{E. Nakar}

\author{D. Maoz}

\affil{School of Physics and Astronomy, Tel Aviv University, 93387 Tel Aviv, Israel}

\author{B. Trakhtenbrot\altaffilmark{2}}

\affil{Department of Particle Physics and Astrophysics, Faculty of Physics, The Weizmann
Institute of Science, Rehovot 76100, Israel}

\author{S. R. Kulkarni}

\author{E. S. Phinney}

\affil{Division of Physics, Astronomy, and Mathematics, California Institute of Technology, 91105 Pasadena, California}

\author{J. Topaz}

\affil{Department of Particle Physics and Astrophysics, Faculty of Physics, The Weizmann
Institute of Science, Rehovot 76100, Israel}

\author{C. Beichman}

\affil{Division of Geophysics and Planetary Science, California Institute of Technology, 91105 Pasadena, California}

\pagebreak
\author{J. Murthy}

\affil{Indian Institute of Astrophysics, Koramangala, Bangalore 560 034, India}

\author{S. P. Worden}

\affil{NASA Ames Research Center, Moffett Field, 94035 California}

\altaffiltext{1}{Kimmel Investigator}
\altaffiltext{2}{Benoziyo Fellow}

\begin{abstract}

The time-variable electromagnetic sky has been well-explored at a wide range of
wavelengths. Numerous high-energy space missions take advantage of the dark $\gamma$-ray
and X-ray sky and utilize very wide field detectors to provide almost continuous
monitoring of the entire celestial sphere. In visible light, new wide-field ground-based
surveys cover wide patches of sky with ever decreasing cadence, progressing
from monthly-weekly time scale surveys to sub-night sampling. In the radio, new
powerful instrumentation offers unprecedented sensitivity over wide fields of view,
with pathfinder experiments for even more ambitious programs underway. In contrast, the
ultra-violet (UV) variable sky is relatively poorly explored, even though it offers exciting
scientific prospects. Here, we review the potential scientific impact of a wide-field
UV survey on the study of explosive and other transient events, as well as known classes of variable
objects, such as active galactic nuclei and variable stars. We quantify our predictions
using a fiducial set of observational parameters which are similar to those envisaged for the proposed
ULTRASAT mission. We show that such a mission would be able to revolutionize our knowledge about
massive star explosions by measuring the early UV emission from hundreds of events, revealing
key physical parameters of the exploding progenitor stars. Such a mission would also detect
the UV emission from many tens of tidal-disruption events of stars by supermassive black holes
at galactic nuclei and enable a measurement of the rate of such events. The overlap of such a
wide-field UV mission with existing and planned gravitational-wave and high-energy neutrino telescopes
makes it especially timely.

\end{abstract}

\section{Introduction}

The coming decade is expected to be a golden age for time-domain
astronomy, which has been identified as an area of unusual discovery
potential by the 2010 Decadal Survey.

There are three reasons for this developing focus. In most electromagnetic
bands the static sky has been imaged to interesting depths:
FIRST \citep{becker95first} and NVSS \citep{condon98nvss} in radio, 2MASS \citep{skrutskie062mass},
UKIDSS \citep{lawrence07ukidss} and WISE \citep{wright2012wise} in IR, SDSS \citep{york2000sdss}
in optical, GALEX AIS in UV \citep{martin2005galex}, and ROSAT \citep{voges99rosat} in X-rays.

Next, technology is now enabling efficient monitoring of large swaths of sky. Advances include
arrays of sensitive detectors in the IR, visible and UV, as well as increase in computing power,
data storage capacity, and improved communications. Following advances in astronomical software designed for
handling large-area static surveys, new development is focussed on various aspects of time-domain investigation, such as variability detection,
in particular via image-subtraction methods \citep{alard98, bramich2008}, event classification \citep{bloom2012, brink2012} and real-time processing and
follow-up \citep{galyam2011}.

Finally, some of the most exciting frontiers, particularly those related to cosmic explosions, require wide-field time-domain
imaging surveys. Examples include the discovery of rare or unusual transient events, \citep{quimby2007,
barbary2009,galyam2009nat, quimby2011,
galyam2012sci, gezari12nat, chornock2013, cenko2012, cenko2013}, as well as systematic studies of unbiased
object samples \citep{arcavi2010, neill2011, vanvelzen2011, quimby2013, gezari13galex}. An additional exciting prospect is the discovery and follow-up of
electromagnetic counterparts of poorly-localized non-electromagnetic signals, such as high-energy neutrino or gravitational-wave
sources \citep{nakar2011nat, metzgerberger2012, ligo2012, abbasi2012}.

It should therefore not come as a surprise that the Large Synoptic Survey Telescope (LSST),
a wide field optical survey telescope, was the top choice of the US astronomical community
for ground-based astronomy. Radio astronomers are developing powerful mapping machines such as LOFAR,
MWA and APERTIF (and ultimately SKA) that will also study transients. The dynamic sky has been a major
science driver for high-energy X- and $\gamma$-ray astronomy, with current (Swift, Fermi, MAXI/ISS) and future
(e.g., AstroSAT, SVOM, LOFT) space missions offering wide-field capabilities.

In contrast, there has been little time domain study and exploration in the Ultra-violet (100-300nm). This is all the more
surprising given that several major questions in astronomy (reviewed below) can be addressed even by a modest UV time domain explorer.
Historically, exploration of the UV by photoelectric missions (OAO 2-3, TD-1A, ANS) provided photometry and spectroscopy of stars and studied bright Galactic variables and novae \citep{code1970, rogerson1973}. Subsequent missions such as IUE, FUSE and HST focused on single-object spectroscopy,
with narrow fields of view prohibiting survey operations.
Only in 2003 the 1.2 deg$^2$ field GALEX mission \citep{martin2005galex} begun the first
systematic study of the extra-galactic static sky, and a more very limited time-domain program \citep{welsh2005, gezari13galex}.

Below we give an overview of possible science that will be enabled by a wide field UV transient explorer. $\S~$\ref{sec:SSB} discusses how the death of massive stars can be explored.
In $\S~$\ref{sec:GRB} we describe how Gamma Ray Burst (GRB) afterglows can be explored and orphan GRBs discovered. In $\S~4$ we
discuss the study of tidal disruption events (TDEs) and in $\S$~5 describes studies of AGN variability.
Sections $\S$~6 and $\S$~7 discuss planetary transits and variable star studies, while $\S$~8 describes
solar-system studies. In $\S$~9 we consider searches for the electromagnetic counterparts to
Gravitational-Wave (GW) and high-energy neutrino sources. In $\S$~\ref{sec:ultra} we describe the
design parameters of the proposed ULTRASAT mission which we use as fiducial to quantify our
predictions above. We conclude in $\S$~11 with a summary of the expected impact of an ULTRASAT-like mission.

\section{Death of Massive Stars}
\label{sec:SSB}

The explosive death of massive stars as SNe is a complex unsolved
astrophysical questions, defined as a science frontier question by the 2010
decadal survey\footnote{New Worlds, New Horizons in Astronomy and Astrophysics, pages 57, 247, SSE-3}. Determining the physical properties of massive stars prior
to explosion is a critical step towards solving this problem; the pre-explosion stellar state sets the
initial conditions to any computational investigation of the explosive process. Direct
identification of SN progenitor stars in pre-explosion images is limited, as it can only be applied to explosions in nearby galaxies (typically $\sim20$\,Mpc away) and requires
that high-spatial-resolution and deep images (mostly by {\it HST} were acquired prior
to the explosion; these conditions are met only for a couple of events per year. Only about
ten such relatively nearby massive stars have been confirmed as SN progenitors \citep{smartt2009, maund2013}.

\begin{figure}
\centering
\includegraphics[width=0.75\textwidth]{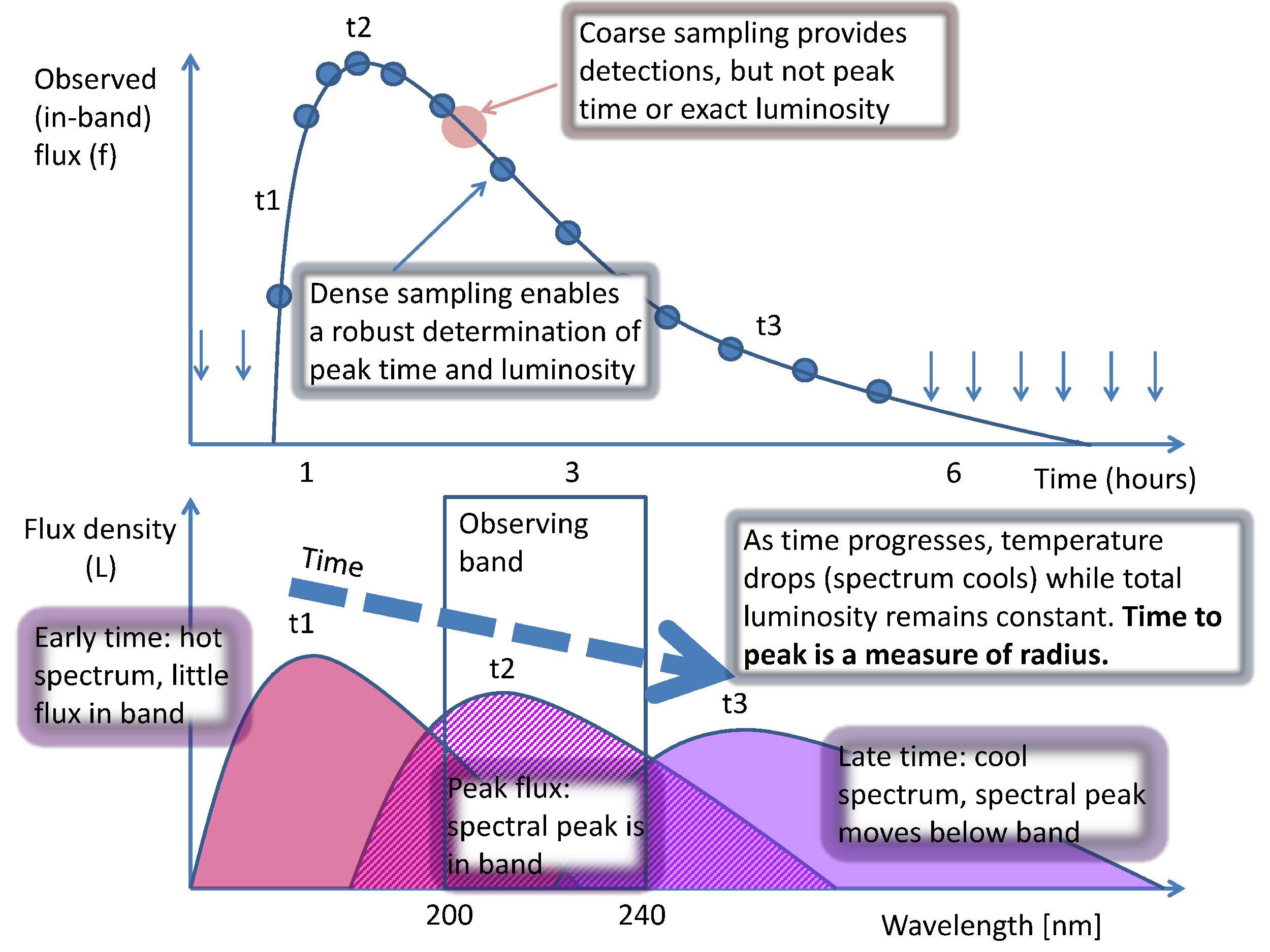}
\caption{Bottom: a heuristic description of shock cooling emission
from a massive star explosion. While the bolometric luminosity is almost
constant, the decline in temperature leads to a rise and fall when observing
in a given UV band. A well measured UV light curve (top) provides a measurement of the
radius and surface composition of the exploding star.}
\label{figshockcooling}
\end{figure}

Early UV observations
of SN explosions provide a powerful method to study the properties
(e.g., radius, surface composition) of exploding massive stars \citep{chevalier1992, matzner_mckee99, nakar_sari2010, rabinak_waxman11}. Following
the supernova explosion a shock wave propagates outward from
the core of the star through its optically thick envelope. When the shock wave
reaches the outer regions where the optical depth is such that the photon diffusion time scale
is shorter than the hydrodynamical time scale, the photons can escape the star; this is
usually called the {\it shock breakout} flare, and would constitute the first electromagnetic
signal for the explosion that an outside observer can detect. For
supergiant stars the initial shock breakout signal is expected to
be in the X-ray/UV and its duration is directly proportional to the radius of the
progenitor star. Following this initial
flare, the thermal energy deposited by the shock in the expanding envelope continues to
diffuse out; we will call this the {\it shock cooling} emission.
The bolometric luminosity of the shock cooling emission is almost
constant, while the temperature of the radiating gas declines.
The shock cooling signal will be prominent in the
UV. The measured flux will rise as the peak of the emitted spectrum
cools and passes though the observed band and will then decline as further cooling
drives the emission peak to redder wavelengths (Figure~\ref{figshockcooling}, bottom). The rate of cooling
(and thus the time it takes for the flux to peak in a given band)
depends on the stellar radius and the composition of the envelope
which determines the opacity. For supergiant star explosions with thick
hydrogen envelopes, the opacity is known (Thomson scattering) and
time independent, so the radius is straightforwardly inferred. For
compact Wolf-Rayet (W-R) stars the opacity is time-dependant and a function of the surface
composition (mass fraction of He, C and O). \citet{rabinak_waxman11}
show that, given a well-sampled light curve, one can infer both the
stellar radius and the surface composition, as well as the ratio of explosion energy
to ejected mass (E/M) and the relative extinction towards an explosion.

The shock-cooling emission
lasts a few hours for compact WR stars \citep{rabinak_waxman11,piro_nakar2012} and approximately
a day for red supergiants \citep{nakar_sari2010,rabinak_waxman11}. This early emission provides constraints
on the progenitor radius and chemical composition, which can not be
derived from later ($<$few days) ground observations,
since by the time the shock cools enough for visible light to be below
the peak emissivity, complicating radiation from other sources (e.g.,
radioactivity, recombination) interferes with these measurements, and the total emission
from shock cooling in compact stars become very faint and difficult
to observe.

\begin{figure}
\centering
\includegraphics[width=0.75\textwidth]{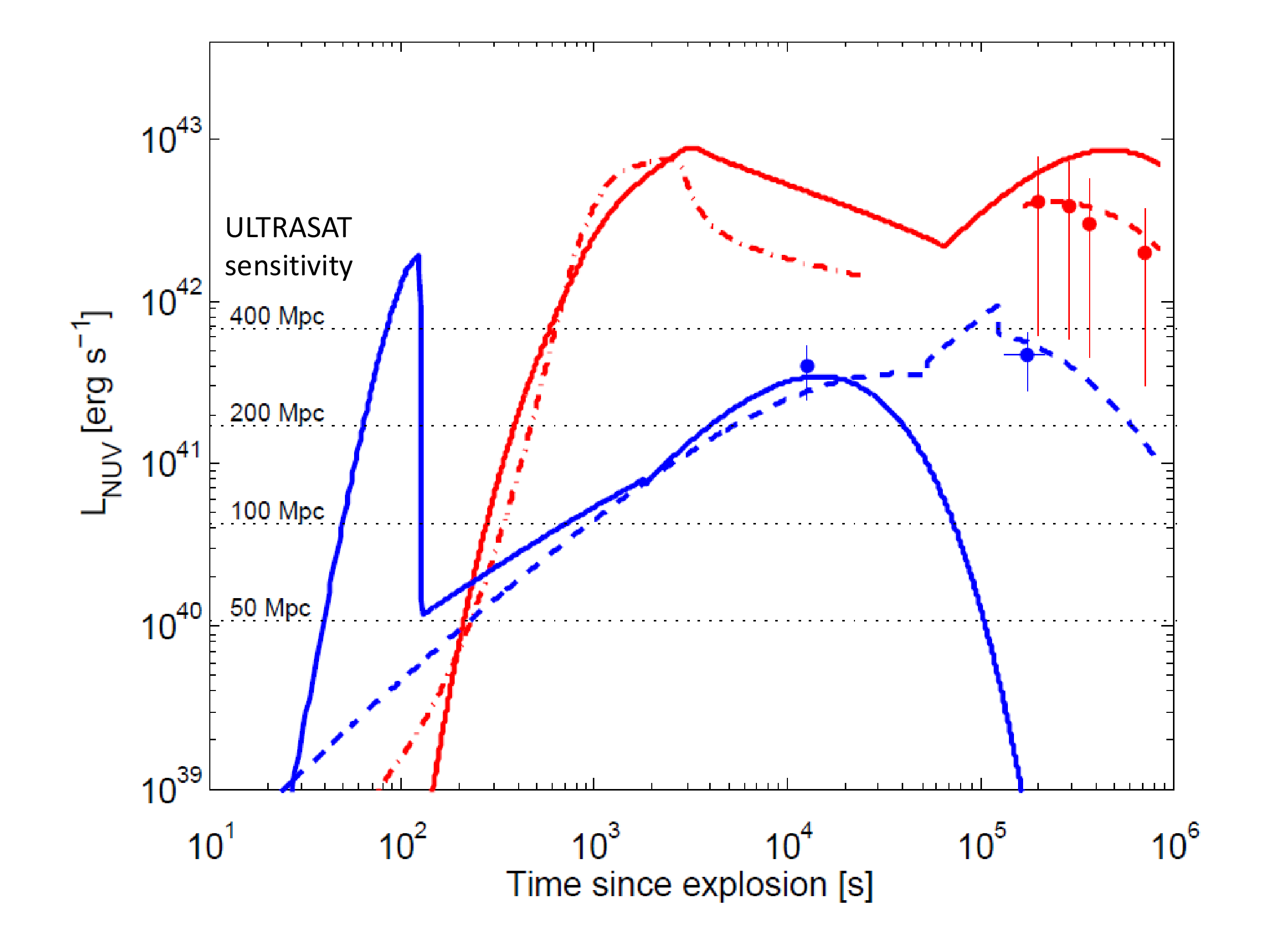}
\caption{Theoretical shock-cooling models in the mid-UV band (200-240nm) (Solid: \citet{nakar_sari2010}; dashed: \citet{rabinak_waxman11}; dash-dot: Sapir et al. in preparation) compare well with
observations (red data points: binned observations of Type II SNe from \citealt{gezari2008sb};
blue points: binned observations of Type Ib SN 2008D from \citealt{soderberg2008}). Red and blue curves
are predictions for red supergiant and W-R star explosions, respectively.}
\label{figshockdata}
\end{figure}

A UV wide-field transient explorer can detect the {\it shock breakout} flare from
the largest stars, and those exploding within an extended circum-stellar medium (e.g.
PTF09uj; \citealt{ofek2010wind}), as well as the {\it shock cooling signal} from
numerous supergiant and W-R massive stars, as predicted by theory and demonstrated by
available observations (Figure~\ref{figshockdata}). Combined space-UV and ground-based
observations triggered by a UV transient explorer would yield an unprecedented wealth of data
about massive star explosions (Table~\ref{tabSNobs}), going beyond the stellar radius
and surface composition (and thus the stellar class of the progenitor: red or blue supergiant, or
W-R star). Such information includes direct measurements of the dust extinction curve
toward the progenitor location, removing uncertainties in measured quantities due to
extinction \citep{rabinak_waxman11} and the amount of radioactive $^{56}$Ni into the ejecta
(a probe of the explosion mechanism and geometry; \citealt{piro_nakar2012}). As the early UV data measure
the ratio of explosion energy to ejected mass, E/M, derivation of the ejecta mass M from
modeling of late-time data (light curves and nebular spectra) would provide information
about the explosion energy. Measurements of early-UV photometry and early spectroscopic
velocity measurements that diverge from predictions of simple models would indicate non-standard
stellar density profiles (e.g., \citealt{bersten2012}) while early UV observations of
explosions occurring in thick circum-stellar medium would probe the final stages of massive star evolution
just prior to explosion (e.g., \citealt{ofek10tel} and references within).
Assuming fiducial survey parameters ($\S$~\ref{sec:ultra}), the known UV signals (Fig.~\ref{figshockdata})
and local SN rates, one could
study $\sim100$ such events per year\footnote{It is interesting to note that should a
nearby SN Ia occur within the surveyed field of view, limits on the shock-cooling emission
from it would place interesting constrains on the progenitor system \citep{piro2010, rabinak2012}}. Analysis of such a sample of massive star explosions would
shed new light on the final stages of massive-star evolution and the explosive deaths of
these stellar giants.

\begin{table}[h]
\begin{tabular}{|lll|}
\hline
Physical property & Required data & Reference\tabularnewline
\hline
\hline
Pre-explosion stellar radius & Early UV photometry & \citealt{nakar_sari2010} \tabularnewline
                             &                     & \citealt{rabinak_waxman11} \\
\hline
Surface chemical Composition & Early UV photometry & \citealt{rabinak_waxman11} \\
\hline
Dust extinction curve        & Early UV photometry & \citealt{rabinak_waxman11} \\
towards the SN               & augmented by ground &                            \\
                             & optical/IR data     &                            \\
\hline
Amount of radioactive        & Early UV photometry & \citealt{piro_nakar2012} \\
$^{56}$Ni mixing into        & plus optical photometry &                          \\
the ejecta                   & and early spectroscopy &                           \\
\hline
Explosion energy             & Early UV photometry    & \citealt{rabinak_waxman11}\\
                             & and late-time ground   & \\
                             & observations           & \\
\hline
Progenitor density profile   & Early UV photometry    & \citealt{bersten2012} \\
                             & augmented by spectroscopic & \\
                             & temperature measurements & \\
\hline
Recent progenitor mass-loss  & Early UV photometry    & \citealt{ofek2010wind, ofek10tel, ofekIIn} \\
                             & combined with spectroscopic & \citealt{chevalier_irwin2011} \\
                             & velocity measurements  & \citealt{balberg_loeb2011}\\
                             &                        & \citealt{svirski2012}\\
\hline
\end{tabular}
\caption{Physical properties derived from early UV observations of massive star explosions}
\label{tabSNobs}
\end{table}

\section{GRB afterglows}
\label{sec:GRB}

\subsection{Untriggered GRBs}

More than half of all Gamma-Ray Bursts (GRB) are associated with an optical/UV counterpart
signal that lasts between minutes to few days \citep{cenko2009}.
The afterglow is supposedly generated by the interaction of relativistic expanding shells with the surrounding medium.
The resulting light curve is a complex, time-dependent combination of
several components (reverse shock, jet break, density bumps, late energy injection) and unraveling them provides
valuable information regarding the physics and energies of the explosion.

Detection of the afterglow in the near UV is limited to the closest GRBs since above $z\sim1$,
intergalactic Lyman absorption will suppress the signal. In addition, UV observations
suffer from higher host galaxy extinction. We estimate
the fraction of GRB afterglows that will be detectable in the UV at a fiducial sensitivity ($\S$~\ref{sec:ultra})
using the complete sample of \citet{cenko2009} and typical UV-R afterglow colors ($UV-R=0$\,mag).
Of the $\sim1000$ GRBs occurring every year, about
10 will happen within the field of view of a mission with fiducial parameters as specified in
$\S$~\ref{sec:ultra} (covering $\sim1.2\%$ of the sky at any given moment, down to NUV$\sim21$\,mag AB), and will be detectable.
These GRBs
will therefore be observed regardless of a high-energy trigger, and their early afterglow
emission will be followed continuously at minute-timescale temporal resolution.

\subsection{Orphan GRB}

Gamma-Ray Bursts are assumed to be collimated explosions, powered by ultra-relativistic jets
that are a few degrees wide (see \citealt{piran2004} for review). Although indirect evidence
supports this model, a direct observational
demonstration of the collimated nature of the outflow would be very
valuable.

A testable prediction of the narrow jet model is that the radiation beaming angle should become
wider with time as the jet decelerates. Thus, low-energy afterglow
emission recorded hours-days after the burst should be seen by observers
out of the initial opening angle of the prompt gamma-ray emission cone.
The hypothesized event of afterglow emission seen without a high-energy (gamma- or X-ray)
emission has been termed ``orphan GRB'' \citep{rhoads1997}.

The orphan events are 100 time more abundant \citep{guetta2005, ghirlanda2013} but are orders of magnitude fainter than the GRB prompt emission and have not been detected so far, with a single possible
exception \citep{cenko2013}. In flux limited surveys we expect a $\sim1:1$ ratio of
orphan to regular GRBs \citep{nakar2003}. A wide field UV transient explorer will thus be able to detect of order 10 events per year. Assuming future high energy missions maintain the current sky coverage
provided by {\it Swift} and Fermi/GBM ($\sim50\%$), we can expect a handful of bona-fide orphan afterglows
per year, i.e., events which are detected as UV transients (with precise temporal information and spatial
localization) and yet have no high-energy detection, even though they have occurred within sky areas covered by sensitive space missions. Later optical/radio observations would be useful to confirm the
identity of such transients, e.g., by identification of an associated GRB-SN, or a long-lived radio
afterglow \citep{cenko2013}. Detection of even a single orphan
afterglow will provide a valuable direct confirmation of the GRB jet
model. The ratio of orphan GRB afterglows (prompt UV without high-energy emission)
to normal events (prompt $\gamma$-ray emission and UV afterglow)
will measure the GRB jet opening angle.
A measurement of the average jet opening angle will allow accurate translation
of observed to isotropic energy, settling the true energy budget of
GRBs.

\section{Tidal Disruption Flares}
When a star passes close enough to a super-massive
black hole (SMBH) with a mass $<10^{8}M_{\odot}$ (for a solar
mass main sequence star) it is shredded by tidal forces. Part of
the stellar material goes into a bound orbit, creating a short-lived
accretion disk around the SMBH, leading to flaring emission in UV-X ray light.
Such tidal disruption events
(TDEs), first considered by \citet{lacy82}, are of extreme interest for the following reasons.

First, they are probing the properties of SMBHs in the center of inactive
galaxies, which are difficult to observe by other means. Second, they probe the
stellar population and dynamics at the vicinity of these black holes,
which determines the rate at which the SMBH is fed. Third, they can provide
a robust estimate of the rate of gravitational waves from extreme mass
ratio inspirals. Finally, we have still a lot to learn about the physical
processes that are involved in the sequence of disruption,
accretion disk build up, the accretion itself and the generation of
various types of outflows, all reflected in the emission profile and
spectrum of such events.

A large sample of observed TDEs would be
extremely useful to tackle the above questions. For example, a study of
the dependence of TDE rates and properties on the galaxy type
can provide us with a completely new angle to study the connection
between galaxy formation and the growth of SMBHs; the rates in particular
are expected to be strongly dependant (and therefore an excellent probe) of
the inner structure of galaxies. Triaxial galactic nuclei will lead to enhanced rates
(by $1-2$ orders of magnitude; \citealt{poon2004}) compared to standard spherical
models \citep{magorrian1999}.

To date there are only a handful of TDE {\it candidates}, observed in gamma-ray
\citep{bloom11tde, burrows11tde, cenko2012}
X-ray \citep{esquej08, cappelluti09, saxton12tde}, optical \citep{vanvelzen2011}
and UV \citep{gezari2006, gezari2008, gezari12nat}.
None of these candidates is a confirmed TDE, mainly due to the difficulty of
ruling out contamination by other object classes (mainly unusual SNe near galactic
nuclei, and flares from AGN that are ``mostly dormant'').
At least in one case the evidence seems strong. PS1-10jh \citep{gezari12nat},
detected by the PS1 ground-based optical survey and studied in the UV by
GALEX, shows spectroscopic signatures of a helium-rich outflow, which
may be interpreted as the disruption of a helium-rich stellar core; the
spectrum does not match any known SN class, and is inconsistent with H-rich
outflows from AGN. However,
The current small and non-uniform sample of candidates does not enables significant
progress regarding the interesting open questions described above.
For that purpose, a well-understood sample of dozens of observed TDEs is needed.

The UV is an optimal band to look for TDEs, since this is where a large
fraction of the accretion luminosity is released. The UV emission
is expected to peak on a time scale of $10-30$ days at a luminosity
of $3\times10^{42}-3\times10^{43}$ erg s$^{-1}$ \citep{strubbe2009, lodato2011}.
Thus a mission with our fiducial parameters
($\S$~\ref{sec:ultra}) will be sensitive to fluxes predicted by these models for emission
during the super-Eddington accretion-driven wind \citep{strubbe2009}
and from the accretion disk \citep{lodato2011} out to $2$\,Gpc and $0.5$\,Gpc,
respectively. Adopting
a TDE volumetric rate based on analytic \citep{magorrian1999, wang2004} and N-body simulations
\citep{brockamp11} estimates of galactic TDE rates, and a local density of
SMBHs from \citet{marconi2004, hopkins2007} of
$4\times10^{-7}$\,Mpc$^{-3}$\,yr$^{-1}$, the mission is expected to detect hundreds of TDEs per year.
An independent consistent estimate is obtained from the detection of a single robust
TDE candidate (PS1-10jh; \citealt{gezari12nat}) during the GALEX time-domain survey (TDS;
\citealt{gezari13galex}). The TDS survey covered $40$ square degrees for 6 months (3 2-month
seasons) and found at least one event, leading to an estimated rate of 1 event per 20 square
degrees per year. Our fiducial mission has a similar sensitivity and cover $\sim700$ square
degrees per year of extragalactic high-latitude sky, leading to an estimated detection rate of
$\sim35$ events per year using the above rate ($>10$ events per year at $90\%$ confidence).
Since this rate is only for helium-rich TDEs similar to PS1-10jh, it is a conservative lower limit
on the total rate which will be dominated by disruptions of common, H-rich stars, and provides
an independent confirmation to the theoretically-derived rate above.

\begin{figure}
\centering
\includegraphics[width=0.75\textwidth]{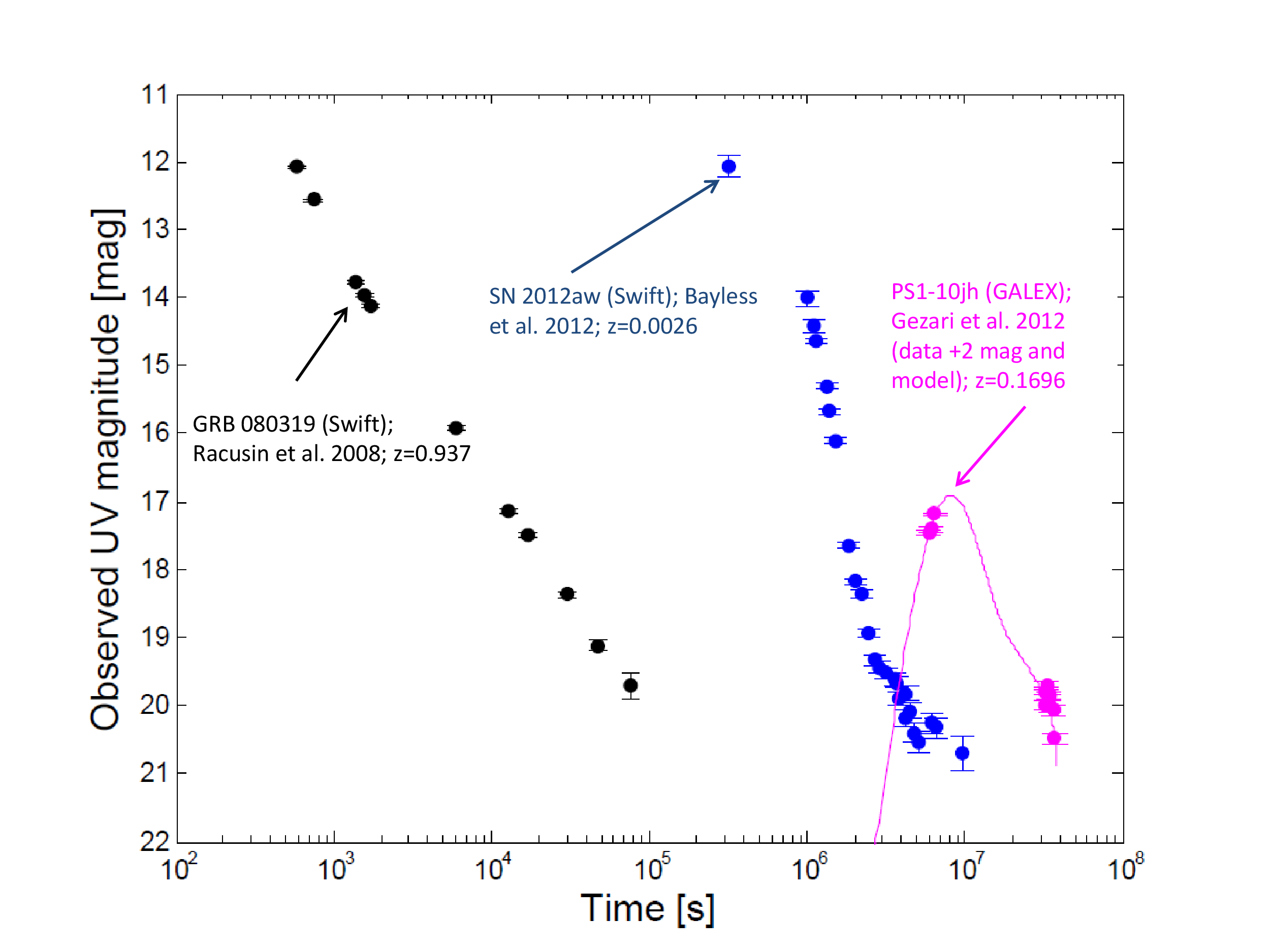}
\caption{UV emission from Supernovae, GRB afterglow and TDEs. Data are
in approximately the mid-UV band (200-240\,nm), and objects would be well
above the sensitivity of our fiducial mission (GRB afterglow data
from \citealt{racusin2008}; SN 2012aw data from \citealt{bayless2012};
TDE data and model from \citealt{gezari12nat}). Object classes are
easily differentiated by their timescales (hours for GRBs, days for SNe, and
months for TDEs). Supernovae will typically occur within relatively
nearby, star-forming galaxies; GRBs occur in much more distant galaxies, while
TDEs typically occur in more luminous hosts, and some would occur in
early-type galaxies with little or no ongoing star formation. Note that a
continuous sky monitoring by a wide-field UV explorer will provide early
UV data, previously only available for GRB afterglows following high-energy triggers,
also for SNe and TDEs.}
\label{figtriplet}
\end{figure}

The UV light curves
will allow a clean separation from other transients (Figure~\ref{figtriplet}).
Detection of such a sample of TDEs will allow a systematic study of their properties
and an examination of the rate as a function of host galaxy. Each detection will also
place an upper limit on the SMBH mass, and thus probe correlations between host galaxy
and SMBH masses (e.g., the M$-\sigma$ relation).

\section{AGN variability}

It has long been hoped that characterizing the multi-wavelength continuum
variability properties of the emission from massive black holes at
the centers of galaxies (Active Galactic Nuclei; AGN) will provide
clues to the physical processes in the inner regions of their accretion
disks, where most of the luminosity is produced. Over the past decade,
this characterization has improved, based both on detailed many-epoch
studies of samples with a few to about a hundred AGNs \citep{giveon99, kelly2009, mushotzky2011}, and on
few-to-tens-of-epoch studies of samples with thousands of objects
(e.g., \citealt{vanden_berk2004, sesar2006, wilhite2008, welsh2011, schmidt2012, macleod2012}).

However, due to the limitations of both approaches, the picture is
still unclear regarding the form and amplitude of the variability
power spectrum, its dependence on physical parameters (redshift, luminosity,
BH mass, Eddington ratio), and the correlations between variations
in different wavelength bands (X-ray, UV, optical). Based on studies
of a handful of Seyfert galaxies with good sampling, the power spectrum
can be described as several broken power laws, with break frequencies
that may scale inversely with BH mass, analogous to results for stellar-mass
accreting BHs. \citet{mushotzky2011} have recently used data for
four Seyfert galaxies in the $\mbox{Kepler}$ field to probe accurately,
for the first time in the optical range, timescales as short as one
month to a day. They found a steepening of the power spectrum to slopes
of $-3$ on these timescales, steeper than ever seen for AGNs either
in the optical or in X-rays.

Figure \ref{fig:AGN1} shows that observing in the NUV probes the inner regions of the accretion disk; NUV-band observations are sensitive to radii which are
almost an order of magnitude smaller compared to those probed by $R$-band data. This makes a UV study complementary to large
optical surveys, and yet opens a window into a relative unexplored region of the accretion
disk around AGN.

\begin{figure}[h]
\centering
\includegraphics[width=0.75\textwidth]{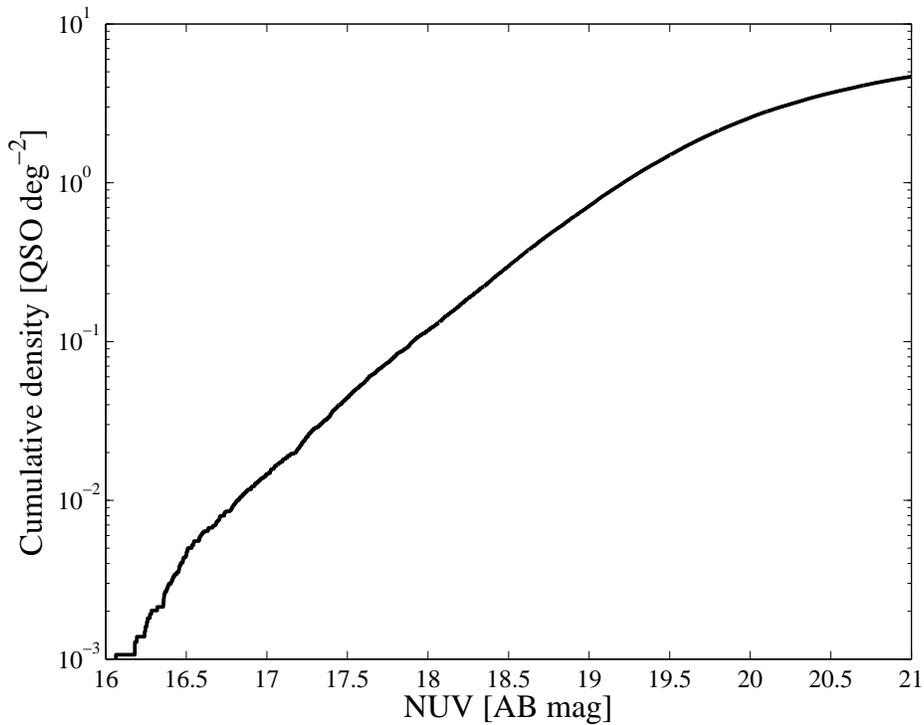}
\caption{The cumulative areal density of spectroscopically confirmed, optically-selected \textit{and} NUV-detected QSOs, as revealed by cross-matching the SDSS/DR7 QSO catalog of Schneider et al. (2010) and the GALEX/GR5 catalog. We only include sources with $\Delta{\rm NUV}<0.3$\,mag. This density is a lower limit on the areal density of QSOs in the NUV band.}
\label{fig:AGN}
\end{figure}

\begin{figure}[h]
\centering
\includegraphics[width=0.75\textwidth]{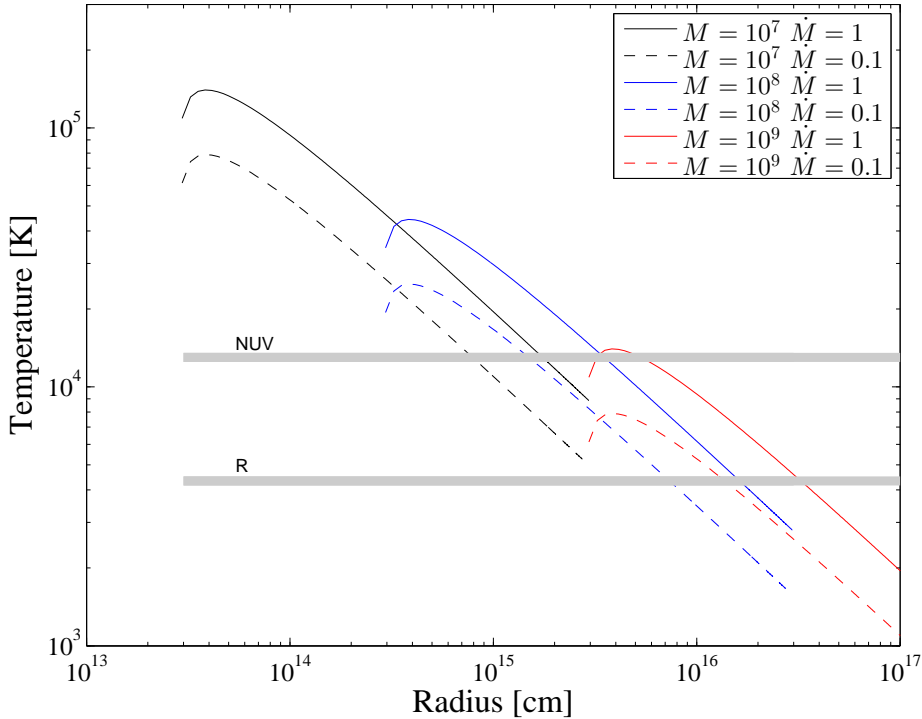}
\caption{The temperature as a function of radius in a simple, optically thick, thin accretion disk for several Black hole mass (marked by $M$ in units
of M$_{\odot}$) and accretion rate (marked by $\dot{M}$ in units of M$_{\odot}$\,yr$^{-1}$) values.
The disk is assumed to span between 10 to 1000 Schwarzschild radii. The gray lines shows the temperature probed by the NUV and $R$ bands. The UV probes significantly smaller radii.}
\label{fig:AGN1}
\end{figure}

A UV transient explorer that would combine a wide field of view with
minute-scale cadence will chart previously unexplored territory in
quasar variability, obtaining both large samples of quasars, and many
epochs per quasar. With a full month of continuous minute-scale exposures
per field, the power-spectrum can be measured down to short timescales
similarly to \citet{mushotzky2011} but, rather than for four AGNs,
for a significantly larger number of quasars (Fig~\ref{fig:AGN}).

Our estimate of the expected areal densities of QSOs for our fiducial survey
parameters ($\S$~\ref{sec:ultra}), are based on the UV properties of a large number
of SDSS QSOs observed in the UV by GALEX. We begin from the SDSS DR7 Quasar Catalog
\citep{Schneider2010}, listing 105,783 spectroscopically confirmed, optically-selected, un-obscured (i.e., broad-line or ``type-I'') AGNs,
most of which are $i < 19.1$\,mag
sources at $z<2$, selected over an effective area
of 9380 ${\rm deg}^2$. We then used the cross-matched GALEX/GR5 - SDSS/DR7 catalog of
\citet{Bianchi2011}, to obtain NUV magnitudes (and errors) for all the sources.
We used data obtained from all GALEX surveys, and limit our analysis to sources
with photometric errors $\Delta{\rm NUV}<0.3$\,mag.

Figure~\ref{fig:AGN} presents the cumulative areal density of SDSS quasars, according to their GALEX NUV fluxes.
At our fiducial flux limit  per single visit (${\rm NUV}=21$), we estimate a density of $>4.5 \, {\rm deg}^{-2}$ sources. Assuming the fiducial field of view and yearly observing cycle of $\sim10$
independent extragalactic footprints, the expected number of AGN surveyed for variability is $\sim36000$.

We note that this is a conservative lower limit, due the combination of several limitations of the data we used. First, the SDSS QSO catalog is naturally flux-limited, and thus does not contain (optically) faint QSOs, which might still be detected in the UV. Second, the GALEX catalog we used (GR5) includes mainly  the rather shallow AIS survey; the fiducial survey we consider here will be significantly deeper once
coadded data are considered, and will thus include more UV-faint AGNs.
In this context, it should be noted that the cumulative areal density of the SDSS QSOs \textit{alone} (not shown here) reaches $\simeq10 \, {\rm deg}^{-2}$ and about .

Such a UV explorer mission would cover the poorly studied UV, and rest-frame far
UV for the higher-z quasars (rather than the optical, where variations
are smaller; e.g. \citealt{giveon99, welsh2011, schmidt2012}).
Finally it will study bona-fide luminous quasars (which
may possibly have lower variation amplitudes than the Seyferts, but
this is basically unknown for these timescales). The sub-percent photometric
precision for measuring comparable variation amplitudes on one-day
timescales (as seen by \citealt{mushotzky2011}) can be achieved by
integrating many $(>100)$ images with lower precision over a day
(and further benefit will be provided by time dilation of high-redshift
sources). The large sample will allow to bin objects according to the
physical attributes derived from their spectra (e.g., black-hole mass and Eddington
ratio) and inspect variability properties within each bin.

A UV transient explorer therefore has the potential for important
new discoveries in the variability at short timescales in the UV for
tens of thousands of luminous AGNs. Results would need to be reproduced by any
successful model of AGN accretion disk physics. Apart from studying
AGN physics, characterizing fast UV AGN variability would be important
for filtering out false transient alerts by other large surveys
searching for transient signals (e.g. \citealt{galyam2002, macleod2012}),
as well as for producing variability-selected AGN samples for cosmological
experiments like BigBOSS \citep{sholl2012}.

\section{Exoplanet transits}
\label{sec:planets}

Recent years have ushered a revolutionary advance in exoplanet detection.
To date, several hundred stars have been confirmed to host orbiting
planets \citep{howard2012}, most detected by their transit light
curve \citep{bord2003phd} in visible light. Discovery of transiting
planets relies on the detection of a deficit in the photon flux during
the eclipse, a signal whose magnitude scales with the square ratio
of the planetary to stellar radii, the eclipse duration, and the observation
time (here assumed to be 30 days per FOV; $\S$~\ref{sec:ultra}). For a significant detection the
flux deficit must exceeds a threshold noise level. A wide-field UV transient explorer
offers opportunities to detect planets. As quantified below, planetary transit detections
are challenging and require precision measurements, so a conservative approach is warranted in
evaluating the potential for such detections.

Several possible subclasses of sources are of particular interest.
First, we consider planets orbiting UV-bright stars, particularly those
orbiting O, B, and A type stars, for which no extensive survey has
been performed \citep{johnson2011},
and whose radiative envelopes are likely to be photometrically
quiet \citep{simon2002}.
However, we note that the exact level of activity of early type stars
is poorly explored.
Since these types of stars are considerably shorter lived than
Sun-like stars, such detections will provide a snapshot of solar systems
in their early stages of formation, and probe planet formation around massive stars.

We estimate the potential results from a mission with our fiducial parameters
($\S$~\ref{sec:ultra}).
We model factors contributing to the noise including background ($0.1$\,
photons\,\,s$^{-1}$\,pixel$^{-1}$), dark current ($0.05$\,electrons\,\,s$^{-1}$\,pixel$^{-1}$), readout
(4 electrons per readout), digitization noise, shot noise (with a Poisson distribution), and
an additional relative accuracy term. The latter accounts for errors induced
by spatial variabilities in the detector
and temporal variabilities
of the starlight. Flat-field errors that vary among observations, variable background sources,
multiple source confusion and associated jitter errors will all contribute uncertainty and reduce
the effective photometric accuracy, particularly needed for faint sources.
Since some of these factors are still unknown, we consider
the number of planets that may be found as a function of this
additional noise.

\begin{figure}
\centering
\includegraphics[width=0.75\textwidth]{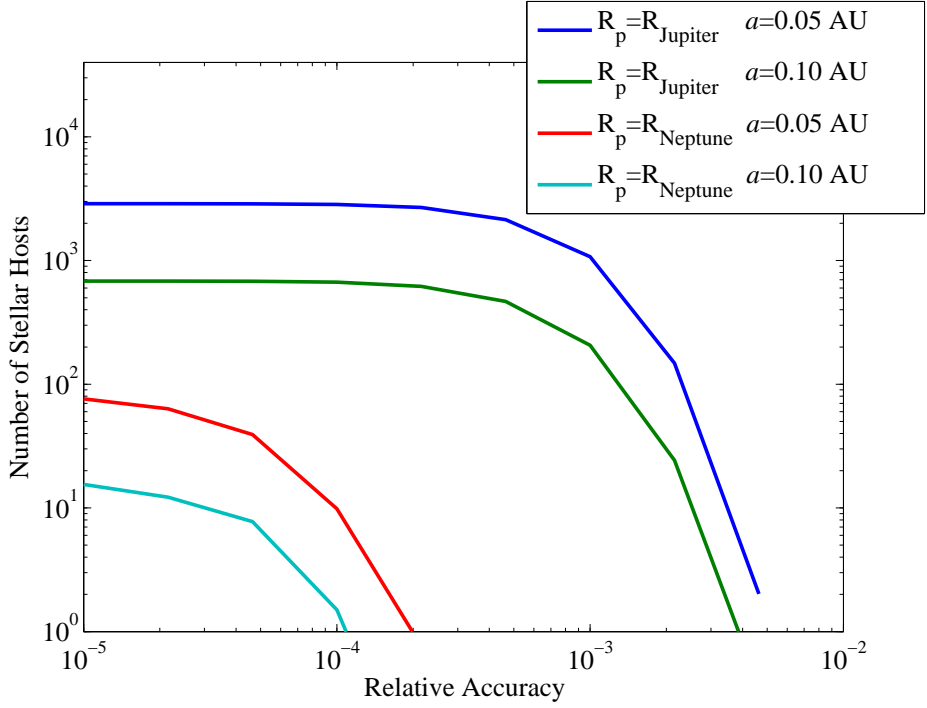}
\caption{We plot the number of stellar sources around which planets would be
detectable by a wide-field UV transient explorer (if they exist),
as a function of the system relative accuracy. Jupiter-size planets would be detectable orbiting
several thousands of stars in tight orbits (a=0.05 AU), assuming an
isotropic distribution of inclinations. This number decreases with
distance from the star, and with planetary radius. Based on experience from
ground-based wide-field experiments using CCD detectors (e.g., PTF; \citealt{ofek2012cal,vaneyken2011,vaneyken2012}),
we estimate a fiducial mission could achieve
a relative accuracy better than $10^{-3}$. With an estimated
hot Jupiter abundance of $1\%$, such a survey will find hundreds of planets
within the mission time.}
\label{fig:planets}
\end{figure}

We estimate the expected number of stellar sources in the
FOV using the Besancon galactic model \citep{robin2003}, corrected
for the Near UV observational band ($200-240\,nm$) using mean spectra of known
stars. We find within each FOV observed for 30 days by the nominal
system there are several thousand stars for which our accuracy is
sufficient to detect a transiting close-in Jupiter-size planet (Figure
\ref{fig:planets}). With an estimated Jupiter-like planet
abundance of 1\% and a relative accuracy term of $10^{-3}$,
the survey will find hundreds
of such planets within the mission time.
It still to be seen if relative accuracy term of $10^{-3}$
is realistic.
Hot stars (O, B, and A type)
constitute a significant fraction of these systems, such that detection (or non-detection),
would constrain the planetary abundance around massive stars, in contrast
with that of the older systems enumerated by Kepler. The Radial Velocity signal of
planets orbiting hot stars is difficult to detect for verification.
However, even a few detections will be important because of the young, relatively
well defined age of such systems.

Secondly, planets transiting evolved stars (giant or sub-giant), whose
photospheric temperatures are significantly lower than those of their
chromospheres, offer a particular opportunity for study
in the UV. Planets orbiting evolved stars were almost exclusively discovered
by the Radial Velocity technique thus far. \citet{assef2009} suggested
a new method to detect planets that orbit such stars, by isolating
the thin ring of chromospheric emission expected near the limb of
evolved stars. The photospheres of these cool stars are significantly colder,
and hence dimmer in the UV. A transit of an evolved star would thus
exhibit two drops in the light curve, as the planet occults the relatively
bright chromospheric ring on the both the ingress and egress parts of
the orbit. Stars with thinner chromospheres would exhibit more distinctive
doublets in their light curves. Extended chromospheres \citep{berio2011}
and those possessing spatial and temporal variability hinder planetary detection,
making the planetary discovery science case less certain.
But the expected occultation signals resolving stellar chromospheres
promise unique and scientifically rich data. In addition to probing planets orbiting
giant stars, that cannot be detected by conventional visible light transits
due to their relative small size, UV observations of such doubly-transiting planets
(and eclipsing binaries in general)
can provide valuable information on chromospheric scale heights of evolved stars.

Thirdly, white dwarfs are rare and small, challenging the detection
of orbiting planets. However they are bright in the UV, and because
of their small size, even relatively small planets can obscure a detectable
fraction of the stellar light. While no such systems have been reported
to date, a large UV sky survey will include more than 20,000 such
objects and may enable the first such discovery (or set a limit on the
planetary abundance around WDs).

\section{Variable Stars}
\label{sec:varstars}
Many types of variable stars have UV amplitudes which are larger than
their visible light variations. Therefore, the UV band may shed new light
on some types of variable stars.
Here we discuss a few of these classes.

\subsection{Eclipsing Binaries}

\begin{figure}
\centering
\includegraphics[width=0.75\textwidth]{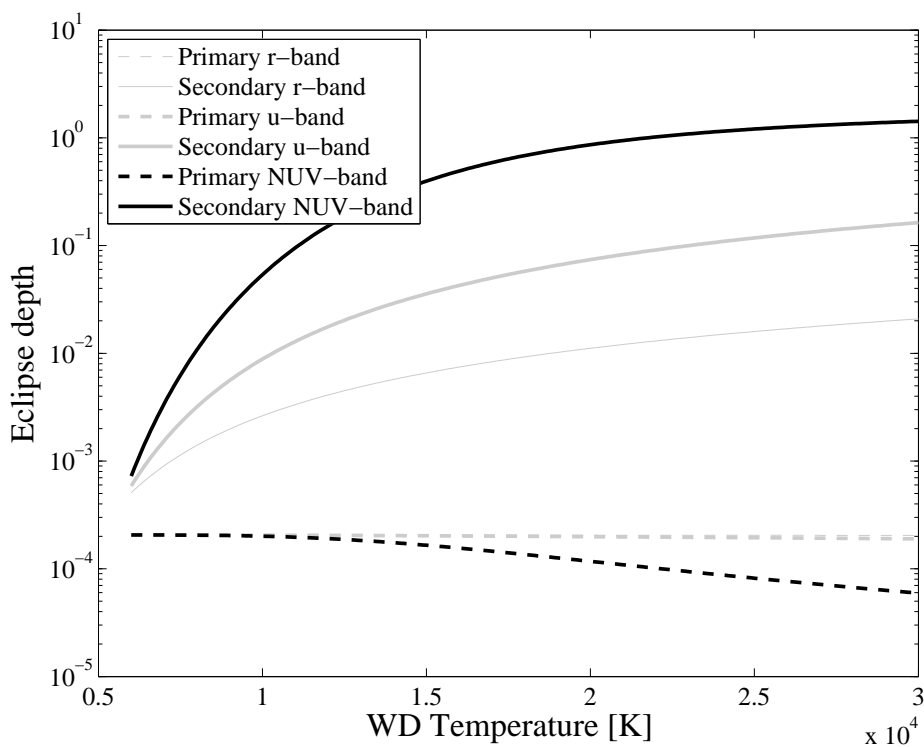}
\caption{The expected primary (solid) and secondary (dashed) eclipse depths for a putative
binary system with a main sequence G dwarf primary and a secondary white dwarf of varying
temperature. One can see that the secondary eclipse is very prominent in the NUV compared to
visible light, more so by
orders of magnitude for hot white dwarf stars. UV observations can thus reveal an important
population of hot WDs in multiple systems with main-sequence stars.}
\label{fig:WD_EB}
\end{figure}

Eclipsing binaries are important for our understanding of stellar parameters
and the distance scale. Observations of these stars in the UV may provide
two important contributions. First, in some cases it may allow us to study
the chromosphere and transition regions of cool stars, via the eclipses
of these regions (see \S\ref{sec:planets}).
For most stars the NUV band is found at the Wein tail of their spectrum,
and the luminosity in this band is very sensitive to temperature
and metallicity. Therefore, NUV observations, along with visible-light observations,
can provide a good description of the components in an eclipsing system, and in
some cases can be used to identify systems that are extremely hard to detect
in visible light.
For example, consider a hot WD eclipsing a main sequence G star. In this case, both primary
secondary eclipse signals will have small amplitudes in visible light. However, the
secondary eclipses will be quite prominent
in the NUV band. Figure~\ref{fig:WD_EB} shows the primary and secondary
eclipse depth as a function of the WD (black-body) temperature, in the $r$, $u$ and $NUV$-bands,
assuming a primary star with a black body spectrum with temperature of 5700\,K and one solar radius,
and a WD radius of $10^{9}$\,cm. One can see that such a system will be difficult to detect in visible
light; but the secondary eclipse signal will be orders of magnitude stronger in the NUV. Such MS-WD
systems (including binaries and triples with two WDs in a tight orbit) are of great interest, e.g., as
putative progenitors of Type Ia SNe \citep{iben1984, katz2012, kushnir2013}.

\subsection{RR Lyr stars}

Having long served as Galactic distance indicators, RR Lyr stars have recently become
a focus of additional scientific interest as popular tracers of Galactic halo
structures (e.g., \citealt{sesar2010}) and are being actively used to search for
Galactic streams (e.g., \citealt{sesar2012}, \citealt{drake2013}).
As the NUV band is in the Wein tail of the spectrum, the NUV variability amplitude of RR Lyr stars
is large compared to the optical variability, making these stars more likely to be detected and
their period easier to measure. In addition,
UV observations of RR Lyr stars present an opportunity to measure their surface temperature
as a function of their phase (e.g., \citealt{wheatley2012}).
Moreover, \citet{wheatley2012} showed that when the effective temperature of RR Lyr stars
is at its minimum, NUV observations can be used to measure the metallicity of these stars.
Such metallicity measurements would be especially valuable for faint and distant
RR Lyr stars that trace ancient Galactic halo structure, providing a new key to
the formation of our Galaxy.

\subsection{Cataclysmic Variables}

Cataclysmic variables are WDs accreting matter from a binary companions \citep{hellier2001}.
Several sub-divisions of CVs are recognized, but a ubiquitous property of this class is
that these systems emit strong episodic flares; these flare are typically very blue, and
could manifest as many-magnitude flares in the blue and UV bands. The combination of
time-scales (days), flare amplitudes (magnitudes) and blue colors of the quiescent counterpart
are quite unique to these systems.

One of the main observational gaps in studies of these systems is the scarcity of
detailed temporal flare profiles; this is due to the fact that the flares are short, and
systems are discovered in flare states by low-cadence surveys; in most cases the flare
is already declining by the time detailed observations begin.

The fiducial mission described in $\S$~\ref{sec:ultra} below would be a great CV discvery machine,
with its very wide field of view and NUV sensitivity, especially during the months that its
orbit drives the field of view across the Galactic plane. The observing mode (almost constant
monitoring) would provide, for the first time, detailed light curves of CV flares at minute time
scales, probably resolving the rise and fine temporal structure. In addition, alerts by such a mission
would allow follow-up studies in other wavelengths (e.g., prompt X-ray studies with {\it Swift} or
a successor mission). Such observations offer a broad potential for progress in deciphering the
complex physics governing the various classes of CVs.

\subsection{Superflares on solar twins and habitability}

The fact that the sun is a relatively quiet star has a benevolent impact
on life on earth (and more recently, on modern technology and space travel).
While many stars in the galaxy are known to episodically emit flares that are
much more powerful (up to $10^{39}$\,erg) compared with the strongest solar flares
($\simlt10^{32}$\,erg; \citealt{schaefer2012} and references therein),
it was often believed that solar twins (i.e., slowly-rotating
G dwarfs) are as quiet as the Sun.

However, it has recently been demonstrated using Kepler data \citep{maehara2012} that
even solar twins have superflares that are orders of magnitude stronger than the strongest
solar flares on record. As these flares have much better contrast in the NUV (flare flux
relative to total photospheric stellar flux) they would be much easier to detect by a mission
similar to our fiducial mission described below ($\S$~\ref{sec:ultra}). A broad census
of the energy distribution and frequency of strong flares as a function of stellar type
and other parameters (e.g. rotation), which such a mission would provide almost by a default,
would be extremely valuable to understand flare physics, as well as any astrobiological
implications. The potential to trigger multi-wavelength follow-up of powerful flares from nearby
solar twins is an especially attractive prospect.

\section{Solar System Objects}

Solar System objects are expected to be relatively faint in the NUV band.
The reasons are that the Solar NUV flux is low and that the albedo of asteroids in
the NUV band is lower by a factor of two relative to their visible light albedo \citep{stern2011}.
Nevertheless, a wide field UV satellite mission will detect a wealth of asteroids and NEOs.
The colors can provide information about their surface composition,
while their magnitude as a function of phase can provide information about their
surface properties (through the opposition effect which is more prominent in the bluer bands;
\citealt{hapke1998}).
Finally, continuous monitoring of asteroids on 10-min to 30-day time scales,
is likely to provide the best sample of asteroid rotation periods (e.g., \citealt{polishook2012}).

Based on a simulations of the known asteroids in the Solar System,
assuming asteroids have a solar-like spectrum and NUV
albedo which is half of the visible light albedo, a fiducial mission ($\S$~\ref{sec:ultra}) can detect
about 60 asteroids in each field of view, of which ~0.5 are NEOs.
Over three years baseline, such a mission can observe over 2600 asteroids
of which 280 are NEOs.

Given the fiducial cadence ($\S$~\ref{sec:ultra}),
we expect to measure the rotation periods for almost all
of these asteroids. Observing solar system objects near opposition,
the mission will generate accurate UV light curves of the opposition effect.
This in turn will allow us to study, in a uniform way, the surface properties of a large sample of asteroids.

\section{Gravitational Waves and High Energy $\nu$ counterparts }

Large scale experiments searching for gravitational waves (GW) from astrophysical
sources, such as Advanced LIGO \citep{harry2010aligo}, Virgo \citep{acernese2005virgo} and
LCGT \citep{koruda2003lcgt} are expected to be operational
in the near future. The best candidates predicted to produce detectable GW signals
are the in-spiral events of double neutron stars
(NS-NS) and neutron star-black hole (NS-BH) binary systems. Correlating
a GW signal from these experiments with an electromagnetic (EM) counterpart
will provide an independent verification of weak signals and therefore
substantially improve the surveys effective sensitivity. In addition, the fine
spatial localization afforded by EM counterparts would provide crucial information
about the sources: measured redshifts of host galaxies would reveal the distance
and thus the energy scale, while environmental information (e.g., young vs. old galaxies)
will shed further light on the physical nature of the emitting systems.

Several possible candidate EM counterparts to GW sources have been identified.
These include Short-duration Gamma Ray Bursts (sGRBs); IR/optical/UV ``kilonova''/``macronova''
\citep{kulkarni2005macronova, metzgerberger2012} and radio afterglows \citep{nakar2011nat}.
GW signal localizations will be, at
best, down to tens of square degrees \citep{nissanke2011}.
Telescopes with exceptionally wide fields of view will be required to search
such large areas. One of the major strength of a wide field space-UV mission would be
its capability to rapidly search and identify
such EM counterparts and pinpoint their astrophysical source, if they are UV luminous.
Both operational directions - a transient detection in the UV of a potential ``macronova''
signal which will be followed
by a search in GW detector data, or a GW signal that will trigger
a target-of-opportunity for the satellite mission - are valuable options.

The estimated GW detection rate, with large uncertainties, is $40$ yr$^{-1}$ for NS-NS
\citep{abadie2010, kopparapu2008} and 10yr$^{-1}$ for NS-BH. Most GW sources that can be detected at reasonable signal-to-noise
ratios would have occurred within $50$\,Mpc, a volume in which
good signal-to-noise ratio in the UV would also be provided by a mission with out fiducial
parameters (down to events $\sim10$ times less luminous than typical SN flares).

A wide field transient explorer may significantly enhance also the sensitivity
of high-energy neutrino detectors \citep{halzen2007,montaruli2012}
such as IceCube by reducing their
backgrounds through precise timing of the SN explosion providing coincidence
with an otherwise sub-threshold neutrino signal. Potential sources include
supernova explosions with failed or ``choked'' jets, that accelerate particles to
high energies (leading to pion and neutrino emission) but fail to puncture the
stellar surface. Such explosions may be quite common (e.g., appearing as normal
SNe Ib/c; \citealt{razzaque2004}). Discovery of even a single case would be
a breakthrough in high-energy neutrino research, as well as in understanding
supernova explosions. Pinpointing an astrophysical source of neutrinos
will shed light also on the related open question of the origin of ultrahigh
energy cosmic-rays \citep{waxman2011}, and will enable probing fundamental neutrino properties
(e.g., flavor oscillations and coupling to gravity) with an accuracy
many orders of magnitude beyond what is currently possible, e.g.,
using upward moving $\tau$ particles \citep{waxman97prl}.

\section{A fiducial mission design: the ULTRASAT mission}\label{sec:ultra}

ULTRASAT - ULtraviolet TRansient Astronomical SATellite - is a proposed wide-field transient explorer satellite mission. The general parameters of the instrument are given in Table \ref{tab:ultra}.

\begin{figure}[h]
\centering
\includegraphics[width=0.9\textwidth]{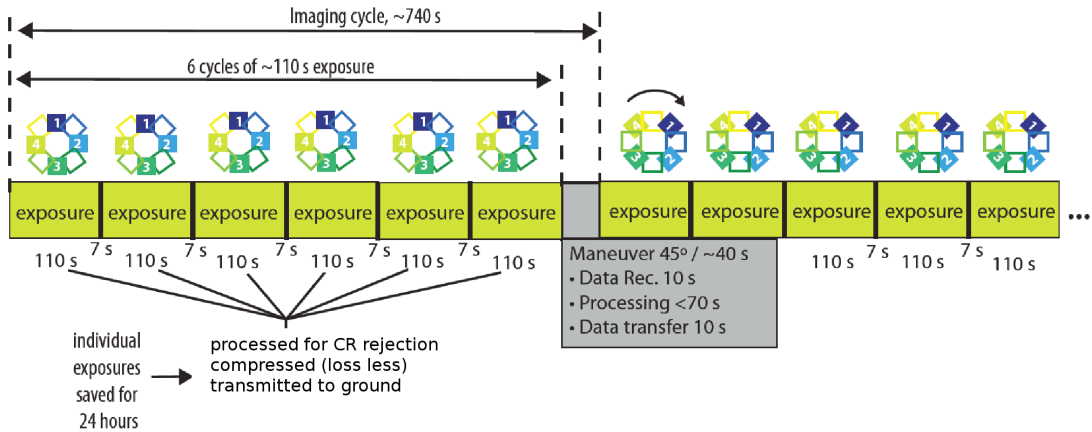}
\caption{The spatial locations of the alternating ULTRASAT field of view.}
\label{fig:fov}
\end{figure}

The instrument includes an array of four identical UV imaging 13.3~cm aperture refractive
telescopes (f2.4), each with a field of view (FOV) of $11^\circ\times11^\circ$.
A UV enhanced $\delta$-doped \citep{nikzad2011} $4k\times4k$ $15\mu m$ pixel
quad-readout CCD will be mounted at the focal plane of each telescope, providing
a pixel scale of 19.3'' per pixel after $2\times2$ binning.
Reflective filters are used to prevent out-of-band light from reaching the CCD detectors.
The spatial resolution is more than sufficient to determine the position of transient sources to within an individual typical galaxy. The telescopes observe perpendicular to the Earth's horizon, and
roughly point always at the anti-sun direction.
The satellite rotates about the FOV axis eight times in an orbit in order to maintain its orientation with respect to the Earth.
The telescopes are arranged such that after a $45^\circ$ rotation alternate fields are interleaved with some overlap, and following an additional $45^\circ$ rotation, the original field is revisited (Figure \ref{fig:fov}).
The total sky coverage is $802$~deg$^2$ (sampled at alternating steps) with $166$~deg$^2$ overlap
observed at every orientation. Each sky position is thus sampled at a cadence between 110s to 24min
during approximately
one month, as the entire field of view drifts across the sky due to the satellite sun-synchronous
orbit.

Each image will be composed of six separate exposures of $\sim110$ seconds stored in memory and co-added after cosmic ray rejection. During each image ($\sim12$~minutes including readout time) the satellite will maintain its pointing and orientation with sub-pixel stability.
The images will then be compressed and transmitted to the ground. In addition, 24 hours of raw individual exposures are stored to provide, on request, higher temporal resolution for specific events (cutouts).
The satellite will be placed at a low Earth orbit with an inclination of $82^\circ$ allowing unobstructed
continuous observations of a region in the direction opposite the sun. As the Earth revolves around the sun the telescope FOV will shift (once a $\sim$~day), keeping it centered at the anti-sun direction and keeping a maximal margin from the Earth, to minimize the effects of stray light and earthshine.
The satellite will be equipped with a real-time communication system based on commercial geostationary communication satellites.
The communication system will have sufficient bandwidth to transmit full images (compressed) to the ground.
The images will be processed within 10 minutes of arrival to identify transient signals and alerts will distributed for follow up by telescopes world wide.
In addition ULTRASAT will has the capability to receive target of opportunity (ToO) alerts. Considering orbit and pointing limitations the telescopes will have the ability to to point at a given ToO within 30~minutes. More details are available at the project website\footnote{http://www.weizmann.ac.il/astrophysics/ultrasat/}.

\begin{table}[h]
\begin{tabular}{|l|l|}
\hline
Field of View & 802 deg$^{2}$\tabularnewline
\hline
Cadence & 12 min\tabularnewline
\hline
Spatial resolution & 19.3''\tabularnewline
\hline
Wavelength band & 200 nm - 240 nm\tabularnewline
\hline
Limiting Mag & 21 ($5\sigma$ detection; 12min coadd)\tabularnewline
\hline
Focal length& $f2.4$\tabularnewline
\hline
Aperture & 13.3 cm\tabularnewline
\hline
\end{tabular}
\caption{ULTRASAT mission design parameters}
\label{tab:ultra}
\end{table}

\section{Conclusions}

We have reviewed the potential science impact of a wide-field UV transient explorer.
In particular, we consider studies of massive star explosions, GRBs and Tidal Disruption
Events (Fig. 2, 3), as well as variability studies of AGN, planetary transits
and various classes of variable stars.
Studies of EM counterparts to GW or high-energy $\nu$ sources also offer an exciting
prospect. We quantify our analysis using fiducial parameters similar to those of the
proposed ULTRASAT mission, and show that such a mission would have a strong impact
e.g., on studies of massive stellar death, GRBs and TDEs, as well as numerous other
subjects including extra-solar planets. It appears that rich scientific returns can
be obtained using a modest space mission focussed on wide-field UV transient surveys.

\section*{Acknowledgements}

We thank M. Van Kerkwijk and B. E. Schaefer for useful discussions. This research has been supported
by grants from the Israeli Space Agency and the Keck Institute for Space Science (KISS).

\newpage
\bibliography{mybib}
\bibliographystyle{apalike}

\end{document}